# One Request, Multiple Experts: LLM Orchestrates Domain Specific Models via Adaptive Task Routing

Xu Yang, Chenhui Lin, *Senior Member, IEEE*, Haotian Liu, *Member, IEEE*, Qi Wang, *Member, IEEE*, Yue Yang, *Member, IEEE*, and Wenchuan Wu, *Fellow, IEEE, Fellow, CSEE*

***Abstract*—With the integration of massive distributed energy resources and the widespread participation of novel market entities, the operation of active distribution networks (ADNs) is progressively evolving into a complex, multi-scenario, and multi-objective problem. Although expert engineers have developed numerous domain specific models (DSMs) to address distinct technical problems, mastering, integrating, and orchestrating these heterogeneous DSMs still entail considerable overhead for ADN operators. Therefore, an intelligent approach is urgently required to unify these DSMs and enable efficient coordination. To address this challenge, this paper proposes the ADN-Agent architecture, which leverages a general large language model (LLM) to coordinate multiple DSMs, enabling adaptive intent recognition, task decomposition, and DSM invocation. Within the ADN-Agent, we design a novel communication mechanism that provides a unified and flexible interface for diverse heterogeneous DSMs. Finally, for specific language-intensive subtasks, we propose an automated training pipeline for fine-tuning small language models, thereby effectively enhancing the overall problem-solving capability of the system. Comprehensive comparisons and ablation experiments validate the efficacy of the proposed method and demonstrate that the ADN-Agent architecture outperforms existing LLM application paradigms.**

## I. Introduction

During the transition towards active distribution networks (ADNs) [1], [2], two salient trends drive this evolution. First, driven by decarbonization targets, the massive integration of distributed energy resources (DERs) [3], [4], introduces significant operational challenges due to their intrinsic uncertainty and volatility [5]. Second, reduced entry barriers in electricity markets [6] facilitate the participation of diverse entities such as prosumers and aggregators [7], complicating real-time management with their heterogeneous objectives and constraints [8]. Together, these trends transform traditional ADN operation into a complex, multi-scenario, and multi-objective problem, necessitating intelligent and proactive management to ensure the overall system safety and efficient coordination [9].

In practice, expert engineers in ADNs have already developed a wide range of domain specific models (DSMs) to tackle distinct technical problems arising from these challenges, such as data tools, simulation tools, optimization tools, and so on [10], [11]. However, although most of the existing DSMs are well-developed, it is still far from trivial for ADN operators to master, integrate, and coordinate them. From the authors' operational experience in real-world ADNs, the primary obstacles mainly stem from two aspects:

1) *Interface heterogeneity.* Since DSMs are usually developed by different vendors to solve different domain specific problems, their interfaces are inherently heterogeneous. Moreover, the coexistence of massive DERs and market entities makes this issue particularly pronounced in modern ADNs. For example, the data tools are typically managed by the marketing department, whereas the optimization tools fall under the purview of the dispatch department. If ADN operators aim to adjust operational strategies of the system, they inevitably need to integrate sources from different departments—a process that is commonly slow and laborious.

2) *Domain expertise dependence.* The design and deployment of DSMs typically require multidisciplinary expertise, encompassing power system analysis, mathematical modeling, and code programming. Also, with the ongoing expansion of dispatch activities and operational scenarios, DSMs are continuously being developed and updated, making it nearly impossible for ADN operators to fully grasp all of them within a limited time. Therefore, there is an urgent need for an automated intelligent analysis and decision-making approach that coordinates existing DSMs to assist ADN operators in grid operations.

In recent years, the rapid development of large language models (LLMs) has emerged as one of the most transformative breakthroughs in the field of artificial intelligence [12], spurring extensive research and applications across numerous vertical domains, such as finance [13], legal services [14], and healthcare [15]. On the one hand, LLMs provide users with intuitive natural language input and output interactions, greatly improving accessibility for non-experts. On the other hand, domain expertise can be embedded into LLMs through various techniques like pretraining, fine-tuning [16], prompting [17],

This work was supported in part by the Beijing Natural Science Foundation under Grant L243003 and the National Science Foundation of China under Grant U24B6009.

X. Yang, C. Lin, H. Liu, and W. Wu (corresponding author) are with the State Key Laboratory of Power Systems, Department of Electrical Engineering, Tsinghua University, Beijing 100084, China (e-mail: yangxuthu@163.com; linchenhui@tsinghua.edu.cn; liuhaotian@tsinghua.edu.cn; wuwench@tsinghua.edu.cn).

Q. Wang is with the Hong Kong Polytechnic University, Hong Kong, China (e-mail: q-yolo.wang@polyu.edu.hk).

Y. Yang is with the State Key Laboratory of High-Efficiency and High-Quality Conversion for Electric Power, Hefei University of Technology, Hefei 230009, China (e-mail: yue.yang@hfut.edu.cn).

few-shot learning [18], retrieval-augmented generation (RAG) [19], and multimodal inputs, shifting the learning burden from users to models and enabling robust and autonomous task execution in specialized applications. Such advantages also offer a viable pathway toward intelligent operation of power systems.

As a result, an increasing number of pioneering studies have begun to explore the feasibility and implementation approaches of applying LLMs to power systems [20]-[30]. For example, researchers in [22], [23] have leveraged the generative capabilities of LLMs to automatically synthesize and augment power grid models or power system operational scenarios. Researchers in [24], [25] have enabled LLM-based power system simulations through a combination of user manuals, few-shot examples, and RAG techniques. In dispatch applications, the researchers have applied LLMs to a variety of scenarios, such as unit commitment [26], energy management [27], and optimal power flow [28]. Finally, some researchers have also combined the LLM with reinforcement learning (RL), in which the LLM analyzes natural language based dispatch requirements or power flow results to provide auxiliary guidance for RL training process [29], [30].

Despite the abundance of existing research, the aforementioned approaches are all customized for a specific scenario or objective, employing an end-to-end LLM application paradigm that cannot accommodate the multi-scenario and multi-objective demands of future ADN operation, which requires adaptive intent recognition, task decomposition, and DSM invocation. Motivated by these requirements, in this paper, we propose the ADN-Agent architecture, which not only provides a user-friendly interaction for ADN operators but also efficiently coordinates heterogeneous DSMs.

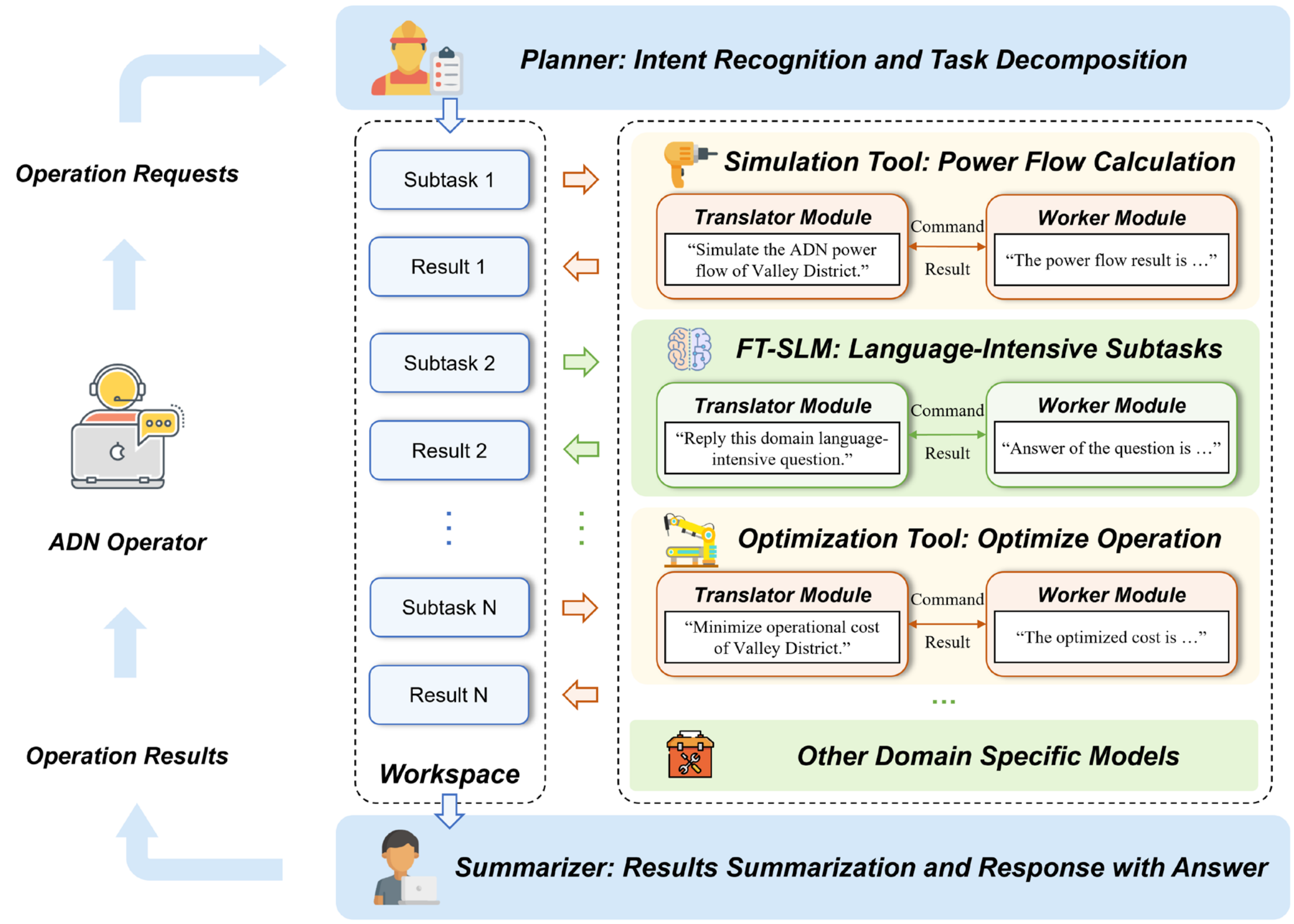

Fig. 1. Overall architecture of the proposed ADN-Agent.

As illustrated in Fig. 1, the proposed ADN-Agent consists of three key components: a general LLM powered Planner, a suite of DSMs designed by expert engineers, and a general LLM powered Summarizer. Upon receiving a natural language operation request from the ADN operator, the ADN-Agent's Planner first analyzes the operator's intent and decomposes this request into a series of manageable subtasks, which are then assigned to appropriate DSMs for execution. Once the DSMs have sequentially completed all subtasks, the Summarizer aggregates and synthesizes corresponding results of the operation request and delivers the final answer back to the ADN operator. This architecture not only enables natural language interaction for ADN operators but also adaptively selects and invokes suitable DSMs based on the requirements of the request, thereby handling diverse problem ranges.

In addition, within the ADN-Agent, in order to coordinate different heterogeneous DSMs, we also propose a novel communication mechanism that provides a unified and flexible interface. Specifically, we first create a dedicated workspace for each operation request, into which all subtask information and execution results are routed, preventing cross-contamination of information across requests. And within each DSM, in addition to the Worker module responsible for concrete subtask execution, there is also a customized Translator module powered by a general LLM, which adaptively retrieves subtasks for the corresponding DSM, translates them into executable commands, and returns their execution results to the workspace. In this way, all heterogeneous DSMs only need to expose their functionalities to the Planner, without providing explicit implementation details. And when invoking a DSM, the Planner merely supplies a natural language description of the subtask. Generation of invocation commands, identification and filling of parameters, and translation of subtask results are all handled by the Translator module, greatly reducing the cognitive burden of the upper-level Planner. Consequently, regardless of the underlying implementation details of individual DSMs, their interfaces to the Planner remain simple, flexible, and unified.

Finally, certain operation requests may involve language-intensive subtasks, such as grid code consultation and ADN model generation, which exceed the domain knowledge present in existing general LLMs. Considering these practical requirements, we innovatively propose fine-tuning a small language model to address this task, and we further develop an automated training pipeline to support it, which encompasses generation prompt design, data generation, data verification, and low-rank adaptation (LoRA) fine-tuning [16]. The fine-tuned small language models (FT-SLMs) significantly outperform general LLMs on domain specific language-intensive subtasks, enhancing the overall problem-solving capability of the ADN-Agent. Contributions of this paper can be summarized as follows:

1) **ADN-Agent architecture for intelligent operation.** To address the multi-scenario, multi-objective nature of modern ADN operations, we propose the ADN-Agent architecture, which comprises a Planner, a suite of DSMs, and a Summarizer. In this architecture, the Planner is responsible for intent recognition and task decomposition, the DSMs execute domain specific subtasks, and the Summarizer compiles and delivers the final answers. By introducing LLMs and leveraging the proposed ADN-Agent architecture, subtasks are dynamically and adaptively assigned by the Planner to the most suitable DSMs according to the operation request, which ensures efficient, flexible, and accurate coordination across varied ADN operational scenarios and objectives, significantly reducing the cognitive burden and scheduling complexity for ADN operators.

2) **Communication mechanism for heterogeneous DSMs.** Given the inherent heterogeneity of existing DSMs, we propose a novel communication mechanism within the ADN-Agent architecture. Specifically, in addition to its original Worker module for concrete subtask execution, each DSM is augmented with a customized Translator module powered by a general LLM, which is tasked with subtask retrieval, generation of executable commands, and result submission. In this way, each DSM only needs to expose its functionality to the upper-level Planner without revealing underlying implementation details, thereby providing a flexible and unified interface. The proposed mechanism not only ensures robust communication and executability of subtasks but also enhances the scalability of ADN-Agent, enabling new DSMs to be seamlessly upgraded or integrated.

3) **Automated training pipeline for FT-SLMs.** Considering that operation requests may involve language-intensive subtasks, we propose fine-tuning small language models and developing an automated training pipeline for FT-SLMs. This pipeline includes generation prompt design, data generation, data verification, and LoRA fine-tuning, integrating complementary strategies for data generation, augmentation, and validation. In the proposed pipeline, domain experts are only required to design generation prompts and perform random sampling-based inspections, significantly reducing their workload. Comprehensive comparisons and ablation experiments validate the effectiveness of the proposed method and demonstrate that ADN-Agent outperforms existing LLM application paradigms.

## II. Preliminaries

### *A. Active Distribution Network Operation*

With the growing integration of DERs and the increasing participation of market entities, ADNs have become significantly more dynamic, complex, and bidirectionally interactive. Under these conditions, traditional passive operation strategies are no longer adequate to ensure reliable and economical grid performance. Instead, proactive and intelligent management is required to respond to rapid fluctuations in renewable energy generation, load levels, and network power flows.

Common ADN operation requests can be broadly categorized into three types: situation awareness, decision-making, and operation analysis. Situation awareness refers to acquiring accurate information of renewable energy generation, load levels, voltage profiles, and power flows to characterize the operational situation of the ADN system. Decision-making involves controlling available DERs and grid equipment to realize certain objectives, such as minimizing cost, minimizing voltage deviation, and minimizing power loss, so that the overall efficiency and security of the ADN can be improved. Operation analysis evaluates how changes in device status, including load variations, equipment switching, new photovoltaic (PV) installations, and topology reconfigurations, affect the operational behavior of the ADN system, supporting more effective planning and adaptive control.

To address these challenges, a wide range of DSMs has been developed by expert engineers. For example, simulation tools can analyze power flow conditions in the ADN and issue alerts for possible voltage violations and branch overloads. Optimization tools can solve a formulated optimization problem to determine optimal control strategies for available DERs and equipment, meeting predefined operational goals. And data tools can retrieve historical data for specific districts and time periods to provide data support. However, leveraging these DSMs requires extensive specialized expertise, and their interfaces are often heterogeneous, making it difficult to master and coordinate them effectively.

## *B. Large Language Model and Domain Specific Model*

In recent years, LLMs have become a cornerstone of progress in the field of artificial intelligence. Built upon the transformer architecture [31], LLMs are characterized by parameter scales exceeding tens or hundreds of billions and are trained on massive text corpora spanning multiple domains and languages. The combination of massive parameter scale and extensive training corpora enables them to internalize intricate linguistic patterns, world knowledge, and reasoning heuristics. Consequently, LLMs exhibit remarkable capabilities in natural language comprehension and generation, robust intent recognition, in-context learning from few-shot demonstrations, and the synthesis of different knowledge into coherent responses, enabling them to assist in addressing a broad range of real-world problems.

Based on these powerful capabilities, LLMs have been increasingly adopted across a range of professional domains, including finance, legal services, and healthcare, where they have demonstrated promising performance in real-world settings. More recently, their application has begun to extend into engineering disciplines, with initial explorations emerging in fields such as power systems, autonomous driving, and industrial process control. These early studies seek to leverage the intrinsic capabilities of LLMs and evaluate their applicability in engineering contexts.

As the research advances, LLMs gradually exhibit inherent limitations in tasks demanding deep domain specific knowledge or precise computation, where relevant information is either absent from their built-in knowledge base or prone to hallucinations and internal errors. To overcome these constraints, recent studies have turned to the integration of DSMs as a pathway to extend their functional boundaries. Among the proposed approaches, two dominant paradigms have been established: one is Function-Call, developed by OpenAI [32], and the other is the multi-LLM collaboration proposed by LangChain [33].

In the Function-Call paradigm, all DSMs are encapsulated as callable functions, exposing their functionalities and parameters to the upper-level LLM. According to the defined functionality of each function, the LLM can dynamically call them during reasoning or content generation process to enhance the accuracy and reliability of its final outputs. For example, researchers in [34], [35] have encapsulated a set of optimization programs as callable functions for the LLM, and the LLM invokes the appropriate function based on the user's intent to generate the corresponding optimized strategy. While in the multi-LLM collaboration paradigm, several LLMs manage different DSM implementations and collaborate to complete complex tasks. This collaborative arrangement effectively reduces the workload on a single LLM and enhances the overall problem-solving ability of the system. For example, researchers in [25] have assigned the power system simulation task to an RAG LLM, a reasoning LLM, and an acting LLM, which jointly solved the problem and improved the accuracy of generated simulation code. And researchers in [27] constructed a natural language to optimization code workflow that decomposes dispatch problems into three stages, i.e., information extraction, problem formulation, and code programming, each handled by a dedicated LLM, thereby improving the dispatch efficiency. In subsequent experiments in Section IV, we will compare the proposed ADN-Agent against Function-Call and multi-LLM collaboration to validate its effectiveness.

# III. Methods

## *A. ADN-Agent Architecture*

As shown in Fig. 1, the proposed ADN-Agent comprises three key components: a Planner, a suite of DSMs designed by expert engineers, and a Summarizer. The Planner is a general LLM that takes natural language based operation requests from ADN operators as input and outputs a set of decomposed subtasks. Its function is to perform intent recognition and analysis of the request, break it down into a series of manageable subtasks, and invoke the appropriate DSMs for execution. Similarly, the Summarizer is also a general LLM that takes execution results from DSMs as input and summarizes the final answer as output. The LLM powered Planner and Summarizer not only provide ADN operators with an intuitive natural language interaction but also enhance the overall intelligence of the ADN-Agent through their inherent reasoning and comprehension capabilities.

As for the suite of DSMs, we integrate the following DSMs into the ADN-Agent to address common problems in ADN operation, including situation awareness, decision-making, and operation analysis:

**Data Tool:** Input is the name of the district and a specified date; output is the PV generation data and load data of the corresponding district on the specified date. The function is to provide data retrieval and analysis for ADN operation.

**Model Tool:** Input is the name of the district, output is the corresponding ADN model of the district. The function is to provide model support for ADN operation.

**Simulation Tool:** Input is the PV generation data, load data, and ADN model; output is the simulation results. The function is to perform power flow calculations on the ADN to analyze its operational status.

**Optimization Tool:** Input is the optimization objective, such as minimizing cost, minimizing voltage deviation, and minimizing power loss; output is the optimized DERs and grid equipment dispatch strategy together with the corresponding objective value. The function is to optimize ADN operation based on dispatch objectives while ensuring safety and meeting the requirements of ADN operators.

**Result Organization Tool:** Input is the simulation or optimization results and an organization request; output is the corresponding organized statistical results, such as the highest voltage, the most congested branch, and the peak load. The function is to distill complex simulation or optimization results into concise statistical ones to provide intuitive feedback for the Summarizer and ADN operators.

Also, it should be noted that one of the main advantages of the proposed ADN-Agent is its high scalability. If a newly developed DSM needs to be introduced, it can be seamlessly integrated into our ADN-Agent following the proposed method.

Based on the architecture of ADN-Agent, in order to enable the general LLM powered Planner to accurately understand the operator's intent and precisely invoke appropriate DSMs, we design the professional prompts tailored for ADN operation. These prompts are meticulously structured into six key components, each serving a specific function to ensure robust and intelligent ADN operation:

1) *Role description*: These prompts position the Planner as an expert in ADN operation, tasked with comprehending operational intents and decomposing them into subtasks.

By explicitly framing the LLM's identity, we effectively align its latent capabilities with the specific domain. This alignment elicits the model's embedded domain knowledge, guiding it to generate professional, context-aware responses rather than generic, superficial suggestions.

2) *ADN operation environment description*: This section equips the Planner with a comprehensive overview of the ADN landscape, detailing common problem types and their typical resolution strategies.

Acting as a "textbook" for the agent, these prompts supplement the external knowledge that the pre-trained model inherently lacks, thereby bridging the gap between general language understanding and specific grid operational requirements.

3) *DSM description*: This component is pivotal for the Planner's ability to orchestrate tasks effectively. It catalogs the available DSMs, delineating their core functionalities and applicable scenarios.

By providing this "technical manual," we enable the Planner to accurately match operational requests with the appropriate DSMs, ensuring that the selected tools are both capable and relevant.

4) *Output format*: These prompts strictly define the interaction protocol, mandating that the LLM generate responses in a standardized JSON format with specific mandatory fields. Additionally, the response must be encapsulated within a designated decorator to ensure clear delimitation and robust automated parsing.

The significance of this constraint extends beyond mere formatting; it facilitates the automated extraction of key information via regular expressions. Furthermore, it also serves as a robust validation mechanism: if the output fails to adhere to the schema, which is often a symptom of hallucination or internal errors, the system can immediately flag the anomaly and trigger a correction loop.

5) *Few-shot examples*: To facilitate in-context learning, these prompts provide curated instances of successful DSM invocations and subtask formulations.

These examples serve as reliable behavioral anchors, implicitly encoding substantial operational experience and domain knowledge. They guide the Planner to mimic the reasoning patterns and decision-making logic demonstrated in the examples, significantly reducing the likelihood of erratic behavior.

6) *Chain-of-Thought (CoT) instructions*: Finally, these prompts impose a structured reasoning pathway, instructing the Planner to sequentially interpret intent, classify the request, and then invoke the necessary DSMs.

This step-by-step cognitive scaffolding forces the model to "think before it acts," which markedly improves the logical consistency and accuracy of the final output.

### *B. Communication Mechanism and Workflow of ADN-Agent*

Given the heterogeneity among DSMs, retaining all execution details imposes a substantial cognitive burden, making it prohibitively difficult for the Planner to directly generate all executable commands. As the number of DSMs increases, the probability of errors caused by the Planner will also rise significantly. In light of this challenge, rather than requiring the Planner to memorize the intricate details of every DSM or generate final executable commands directly, we augment each DSM with a customized Translator module. The Translator module is also powered by a general LLM, which takes the assigned subtask as input and outputs the corresponding executable command. This specific Translator module is responsible for subtask retrieval, the generation of executable commands, and the submission of results. The generated command is then passed to the Worker module within the DSM, such as an optimization solver or a simulation software, which performs the concrete execution of the command. The design of this novel communication mechanism ensures that each DSM only needs to expose its functionality to the Planner, and the Planner can achieve efficient task resolution and DSM orchestration by describing subtasks in simple natural language.

Compared to the Function-Call paradigm, the upper-level Planner in ADN-Agent only needs to perform task decomposition. The detailed generation of concrete commands, execution of functions, and filling of parameters are all handled internally within each DSM by the Translator module and the Worker module, thereby reducing the cognitive burden on the Planner. While compared to the multi-LLM collaboration paradigm, the ADN-Agent benefits from centralized orchestration by the upper-level Planner, which renders inter-module coordination more logical and coherent. Furthermore, the incorporation of DSMs substantially enhances the domain expertise of the underlying LLM. Comparison studies in Section IV will demonstrate the effectiveness and advantages of the proposed ADN-Agent.

To help readers better understand the proposed communication mechanism and the ADN-Agent workflow, we use one simple operation request as an example and illustrate its complete execution process in Fig. 2. When the ADN-Agent receives a request from the operator such as "What is the peak voltage recorded in the Valley District on October 12, 2024?" the Planner first analyzes the request to identify which DSMs are needed and then decomposes it into a set of subtasks placed in the workspace, such as model extraction, data extraction, power flow simulation, and peak voltage computation.

These subtasks are then processed sequentially by their corresponding DSMs. When a subtask is reached, the Translator module within that DSM interprets it and translates it into an executable command. Taking the request illustrated in Fig. 2 as an example, the commands generated by the Translators are implemented as various Python functions. These commands are then passed to their corresponding Worker modules for concrete execution. After the Worker module finishes execution, the Translator sends the result back to the workspace. Once all of the subtasks have been completed, the Summarizer retrieves the execution records from the workspace, synthesizes them into a final response, and returns it to the ADN operator, thereby concluding the overall task. As shown in Fig. 2, the Summarizer outputs "The peak voltage in the Valley District on October 12, 2024, is 1.060 p.u." which directly and accurately resolves the operator's question.

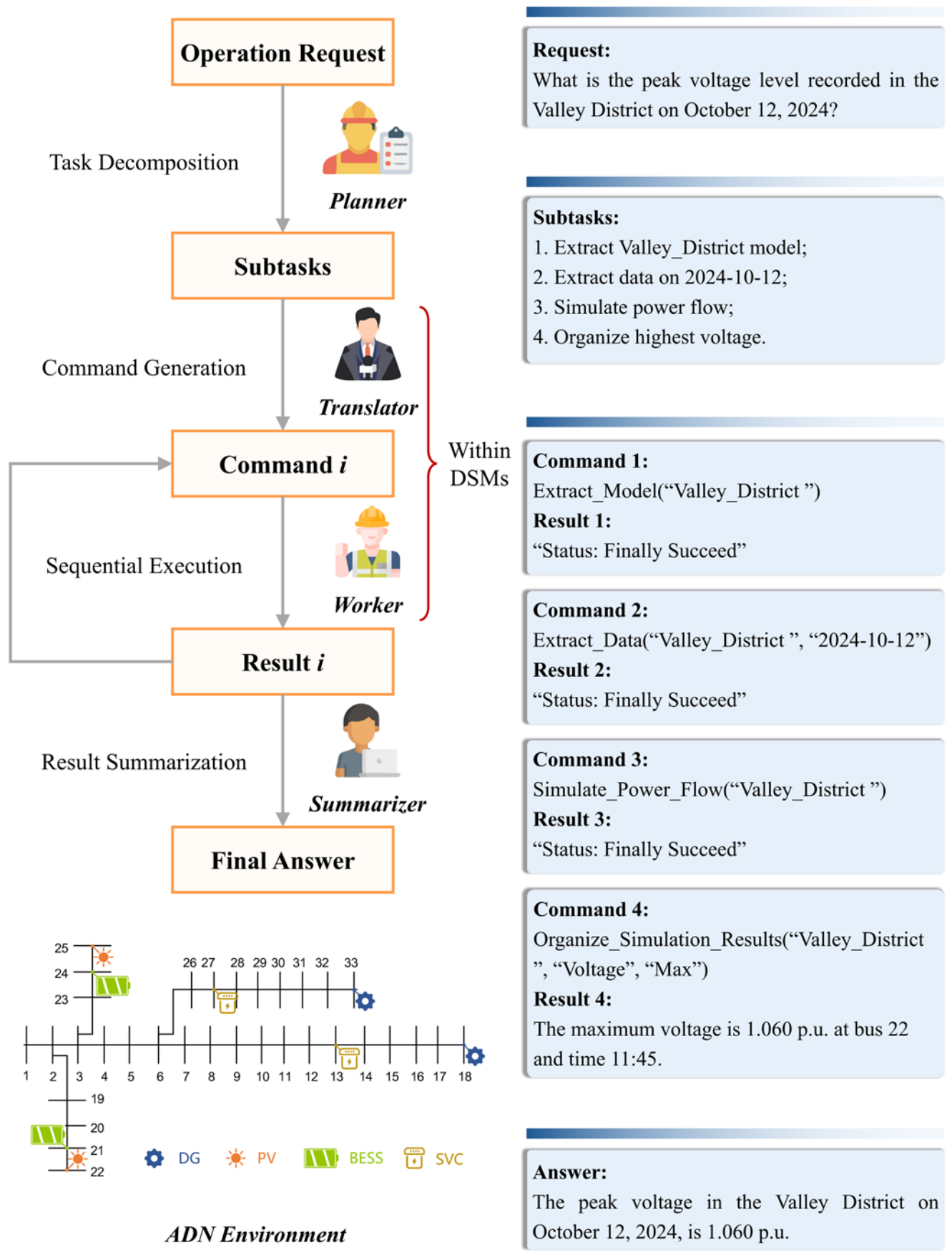

Fig. 2. Illustration of the workflow of ADN-Agent.

With the composition and processing workflow of the ADN-Agent established, we now analyze the rationale and advantages behind its design. First, the separation of tasks reduces the cognitive load on the Planner, thereby decreasing the likelihood of hallucinations or information loss. This design enforces a decoupling between the upper-level semantic comprehension (the responsibility of the Planner) and lower-level execution details (the domain of the Translator and Worker), which prevents the Planner from suffering performance degradation or hallucinations caused by the burden of memorizing complex DSM execution details.

Second, the modular architecture provides enhanced generalization and transfer capabilities, thereby improving the overall scalability of the system. When a DSM needs to be updated, for instance, due to device commissioning/outage, or parameter adjustments, it is only necessary to replace the internal Translator and Worker, without modifying the upper-level Planner. Furthermore, if a newly developed DSM is required to be added, simply informing the Planner of its capabilities enables correct invocation. This design holds significant engineering value in dynamic environments, particularly in modern power systems, where both the operational landscape and DSM technologies are evolving rapidly. Also, drawing an analogy to the real-world scenario, our Planner functions like an experienced operator who, after training, can rapidly adapt to any ADN in any region and under any circumstances to make reasonable decisions. The only requirement is the timely updating of technical manuals (the description of the DSM in the prompts), while the specific maintenance of the DSM is delegated to the underlying engineers (the Translator and Worker).

Third, the introduction of the Translator module also brings about an improvement in security. If we completely rely on a single LLM, the Planner would be likely to generate erroneous instructions. However, the Translator module serves not only as an intermediate bridge but also as a security gateway. Research in the field of LLM suggests that employing multiple distinct LLMs to perform tasks collaboratively can effectively mitigate the generation of errors. Even if the Planner produces minor hallucinations, the Translator module can correct them via pre-defined DSM invocation templates and verification logic, thereby averting potential risks.

And finally, thanks to the natural language inputs and outputs of the Translator, the interpretability and maintainability of the ADN-Agent are also enhanced. The inputs and outputs of the Translator module serve as the intermediate states of the entire ADN-Agent execution. In the form of natural language, they assist the ADN operator in understanding the task execution status and intermediate processes, allowing for timely intervention in case of any possible errors. When the ADN-Agent requires testing or maintenance, the Translator module can also quickly identify the DSMs that need to be updated.

### C. *Training Pipeline for Small Language Models*

As previously mentioned, ADN operation requests may involve language-intensive subtasks such as inquiries regarding grid codes, generation of ADN models, and creation of operation tickets. Since these tasks exceed the capabilities of general LLMs, we propose fine-tuning a small language model to transform it into a specialized FT-SLM dedicated to handling these subtasks. During the ADN operation, model fine-tuning serves three primary purposes:

First, knowledge infusion. Since language-intensive subtasks demand extensive domain expertise that is often scarce in the pretraining corpora of general language models, fine-tuning is necessary to fill this knowledge gap. Second, instruction following. In the ADN-Agent architecture, the DSM carries out Planner-assigned subtasks based on the Translator instructions, requiring fine-tuning to improve the FT-SLM's ability to follow these instructions. Finally, format alignment. In practical ADN environments, the operators may use internal programming languages, entity specific data formats, and proprietary ADN models; therefore, another important role of fine-tuning is format alignment, ensuring that the FT-SLM's outputs meet actual operational requirements.

Conventional model fine-tuning typically relies on human-annotated data, which is often arduous, time-consuming, and susceptible to subjective bias. Moreover, effective mechanisms for verifying and correcting errors introduced during annotation are often lacking. To address this problem, we propose an automated training pipeline for FT-SLMs, whose workflow is illustrated in Fig. 3.

The first stage of the pipeline is generation prompt design, in which domain experts craft prompts that specify the target scenario, define the data generation methodology, and include some few-shot examples. These prompts are then provided to a general LLM for data generation. In this stage, in addition to leveraging the general LLM's ability to generate instruction-answer pairs automatically, traditional data augmentation techniques such as parameter perturbation and synonym replacement can also be applied to further expand the dataset size and diversity. The few-shot examples created by domain experts in the previous stage can also be regarded as high-quality data samples.

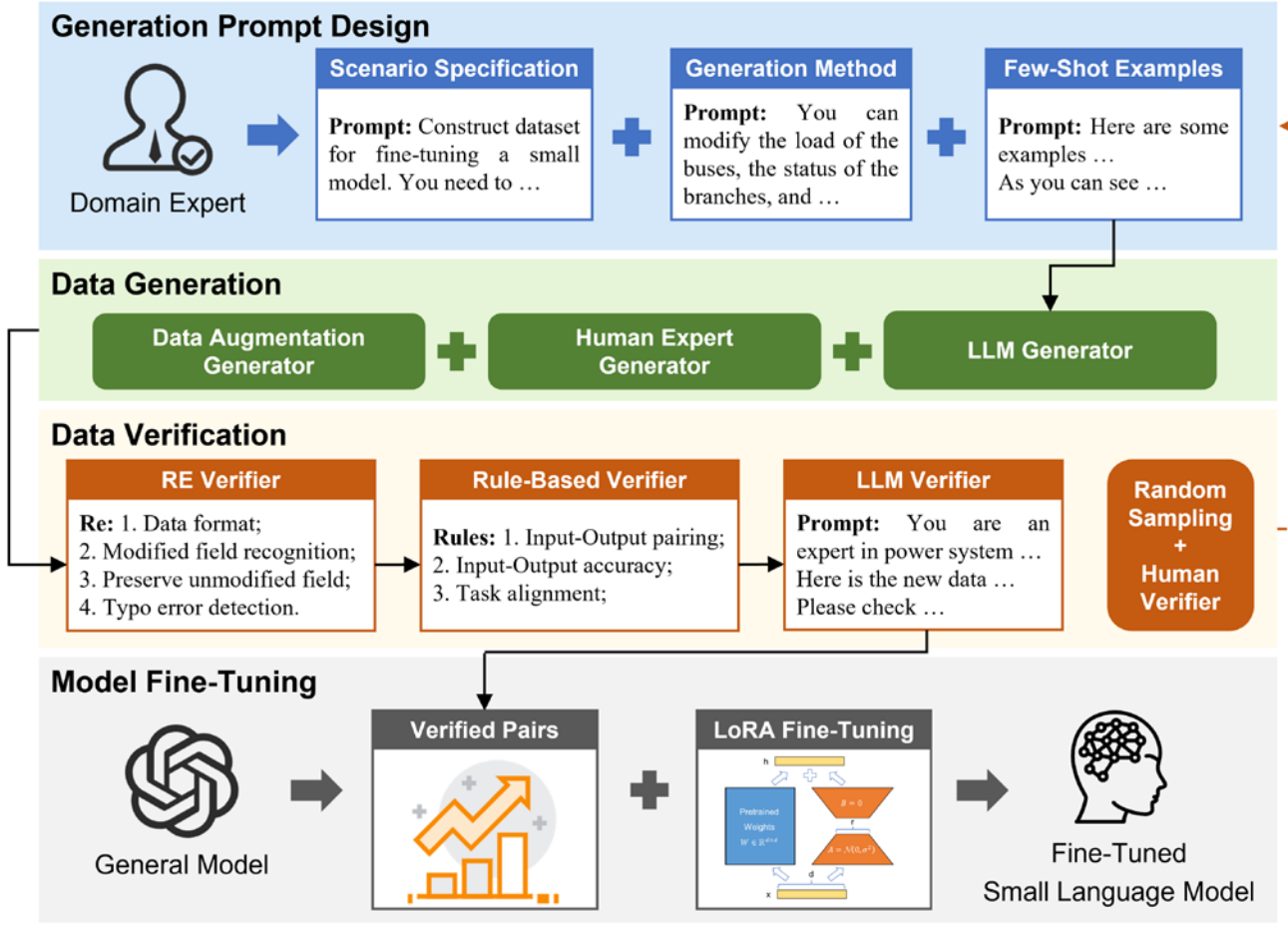

Fig. 3. Automated training pipeline for FT-SLMs.

Each generated instruction-answer pair will undergo the data verification stage, during which we employ a combination of complementary validation methods, including a regular expression verifier to filter out natural language errors, a rule-based verifier to identify inconsistencies within the instruction-answer pairs, and an LLM verifier to detect errors inherent to the LLM generator. Some recent research indicates that employing different LLMs for generation and verification significantly lowers the probability of producing inaccurate or erroneous content [36]. These three proposed verification steps can filter out the vast majority of incorrect data, and domain experts may further conduct random sampling-based inspections. Any errors identified by domain experts can be addressed by refining the generation prompts, thereby enabling an iterative improvement process that enhances both data quality and distribution diversity.

Finally, all verified samples form the fine-tuning dataset. A general small language model is then fine-tuned using LoRA to produce an FT-SLM that is specialized for certain types of language-intensive subtasks.

As can be seen in Fig. 3, throughout the entire pipeline, domain experts are only responsible for generation prompt design and sampling-based inspection of generated data, while all remaining steps are fully automated, substantially reducing their workload. To summarize, this pipeline fundamentally enables domain experts to convey their knowledge to a general LLM through contextual prompts. The LLM then reformulates this knowledge into instruction-answer pairs, which are subsequently used to fine-tune the FT-SLM. Finally, through fine-tuning, this knowledge is internalized within the FT-SLM, thereby achieving a full transfer of domain knowledge, i.e., from prompts to instruction-answer pairs and ultimately into the FT-SLM's parameterized form.

Beyond significantly reducing the manual burden, another major advantage of the proposed training pipeline is its ability to enhance the diversity of generated samples, thereby improving the robustness of the FT-SLM. If the sample distribution during the generation process is found to be too concentrated, the domain experts can directly modify the few-shot examples or the generation prompts. This allows for the iterative improvement of both the quality and diversity of the generated samples. Furthermore, our pipeline employs different LLMs for sample generation and verification, while the FT-SLM serves as another distinct model. This approach of distilling knowledge from heterogeneous models is supported by research in the LLM field and helps mitigate distribution concentration and error propagation to a certain extent.

Simultaneously, the integration of the Translator module within the ADN-Agent architecture serves a critical function beyond its primary role. It ensures that the input fed to the FT-SLM during the inference phase remains as consistent as possible with its training data. Through the processing of the Translator, the FT-SLM receives inputs that are consistent in form with its training distribution, with all required information effectively organized. This significantly improves the model's accuracy. In this way, we effectively avoid the performance degradation caused by out-of-distribution issues, spanning from offline data generation to online inference, which is also validated by our subsequent experimental results.

Specifically, during the ADN operation process, a typical language-intensive subtask is adjusting the ADN model. Because ADNs undergo frequent load variations, equipment switching, new PV installations, and topology reconfigurations, the grid model must be modified or adjusted accordingly to assess the impact of these changes. This subtask requires substantial domain expertise, and its outputs must adhere to the standardized format of power grid cases, making fine-tuning a small language model a suitable solution.

Therefore, for this "model adjustment" subtask, we generate 1680 instruction-answer pairs based on the proposed pipeline, whose input is the original ADN model and an adjustment request, output is the corresponding adjusted ADN model, and its case format is in accordance with MATPOWER [10]. Using these generated pairs, we fine-tune a small language model qwen3-8b into an FT-SLM and encapsulate it as another DSM for our ADN-Agent:

**Model Adjustment FT-SLM:** Input is the original ADN model and an adjustment request; output is the corresponding adjusted ADN model. The function is to adjust the ADN model, and it can be invoked in response to operation requests that involve adjustments to the ADN operating mode.

To conserve space and avoid redundancy, we have included all prompts, including those for the Planner, Translator modules, Summarizer, and FT-SLM training in an online supplementary file for readers' reference [37]. All DSM implementation details are also provided comprehensively.

## IV. Numerical Studies

### A. Methods and Cases Setup

In order to validate the effectiveness and advantages of the proposed ADN-Agent architecture, communication mechanism, and training pipeline, we set up an ADN operation scenario in which the ADN operator manages three districts: the valley district, the railway district, and the business district, which are represented by IEEE 33-bus [38], 69-bus [39], and 141-bus [40] distribution systems, respectively. Controllable devices include micro gas turbines, PVs, energy storage systems, and static VAR compensators.

To evaluate the proposed method and baseline methods, we manually design a benchmark of 40 operation requests grounded in real-world scenarios and frequently encountered problems. This benchmark includes 10 situation awareness requests, 10 decision-making requests, and 20 operation analysis requests. And the covered scenarios involve load variations, equipment switching, new PV installations, and topology reconfigurations. Considering the randomness in LLM content generation, each of these 40 operation requests is evaluated 3 times using different random seeds, yielding a total of 120 runs. The results presented below represent the average value across 120 trials. Detailed network topologies, PV profiles, equipment configurations, and the benchmark are also provided in the supplementary file [37].

Our designed baseline methods consist of three groups: First, in order to validate the efficacy of the proposed ADN-Agent architecture itself, we compare ADN-Agent against two prevailing DSM coordination paradigms: Function-Call and multi-LLM collaboration. In the Function-Call method, all DSMs are encapsulated as callable functions and invoked by a single LLM. In contrast, the multi-LLM collaboration paradigm does not provide pre-encapsulated DSMs. Instead, each LLM is given access to the implementation code of the individual DSMs and is required to coordinate with other LLMs to jointly generate a final executable script.

Second, to evaluate the contribution of the proposed communication mechanism and training pipeline for FT-SLM, we design two ablation baselines called "No-Trans" and "No-FT". In No-Trans, the Translator module of the FT-SLM is removed, such that raw subtasks are fed directly into the FT-SLM without any preprocessing, allowing us to observe the resulting impact on its behavior. And in No-FT, the FT-SLM is replaced with a general LLM to examine the performance disparity between a general LLM and the FT-SLM on the domain specific subtasks, as well as the resulting impact on the overall ADN-Agent. And finally, the "Zero-Shot" ablation baseline is also implemented to assess the proposed prompts, in which all of the few-shot examples in the LLM prompts are removed.

To quantitatively assess the performance of each method, we define three evaluation metrics accordingly, i.e., completion rate, DSM usage accuracy, and result accuracy. Completion rate measures whether a method can successfully produce a response to a given request, thereby reflecting its task executability, i.e., the ability to generate a coherent and actionable output rather than failing or abstaining. DSM usage accuracy evaluates whether the method selects appropriate DSMs and provides accurate subtask descriptions or correct parameter specifications. This metric captures the method's functional comprehension of DSM capabilities, indicating how well it understands what each DSM does and how to coordinate it properly. Result accuracy assesses the correctness of the final output against ground-truth solutions, thereby reflecting the method's problem-solving fidelity, i.e., the capacity to orchestrate perception, reasoning, and DSM use to arrive at a semantically and operationally valid conclusion.

TABLE I
Configurations of Used General Large Language Models

| Parameter | Value |
|---|---|
| Version (qwen-plus) | "2025-09-11" |
| Version (deepseek) | "deepseek-v3" |
| Top-p (qwen-plus) | 0.8 |
| Top-p (deepseek) | 0.6 |
| Temperature | 0.7 |

In the following experiments, the general LLMs used for evaluation include qwen-plus and deepseek, whose configurations are provided in Table I.

### B. Performance of the Fine-Tuned Small Language Model

Before proceeding with the comparative evaluation, we first examine the performance of the FT-SLM on language-intensive subtask in this subsection. In this paper, the FT-SLM is fine-tuned specifically for the subtask of adjusting ADN model. The fine-tuning parameters are listed in Table II, and the performance of each language model on this subtask is presented in Fig. 4.

In Fig. 4, qwen3-8b denotes the base model without fine-tuning, whereas qwen3-8b-ft (No Trans) refers to the fine-tuned variant that receives raw subtasks as input without translation. In contrast, qwen3-8b-ft processes canonical inputs processed by the Translator module. The performance of two general LLMs on this subtask is also evaluated to demonstrate the necessity of specific fine-tuning. Additionally, the "format accuracy" reported in Fig. 4 refers to the proportion of language model outputs that comply with the required structural format. Specifically, in this paper, the output must conform to a valid and executable MATPOWER grid case. While the "result accuracy" measures whether the generated output is semantically correct, i.e., whether the tested language model accurately adjusts the original ADN model in accordance with the subtask specification.

TABLE II
CONFIGURATIONS OF FINE-TUNING

| Parameter | Value |
|---|---|
| Data Generation LLM | qwen-plus |
| Data Verification LLM | qwen-max |
| Fine-Tuning Base Model | qwen3-8b |
| Number of Fine-Tuning Samples | 1680 |
| Batch Size | 16 |
| Learning Rate | 3.0e-4 |
| Learning Rate Scheduler | Linear |
| LoRA Alpha | 16 |
| LoRA Rank | 8 |

As can be seen in Fig. 4, the base model qwen3-8b achieves notably low performance on both format accuracy and result accuracy. This limitation is largely attributable to its relatively small parameter scale, which is insufficient to support complex comprehension and reasoning processes. Furthermore, its pretraining corpus lacks coverage of domain specific knowledge, further degrading its effectiveness on this specialized subtask. In contrast, after fine-tuning, qwen3-8b-ft (No Trans) exhibits a substantial performance improvement, with format accuracy reaching 100%, thereby providing empirical validation of the effectiveness of our proposed training pipeline. Moreover, when the Translator module is incorporated, the result accuracy of qwen3-8b-ft further improves from 88% to 98%. This gain stems from the fact that the Translator standardizes the input into a canonical form that aligns with the input distribution of the fine-tuning data, thereby enabling the FT-SLM to produce more accurate outputs.

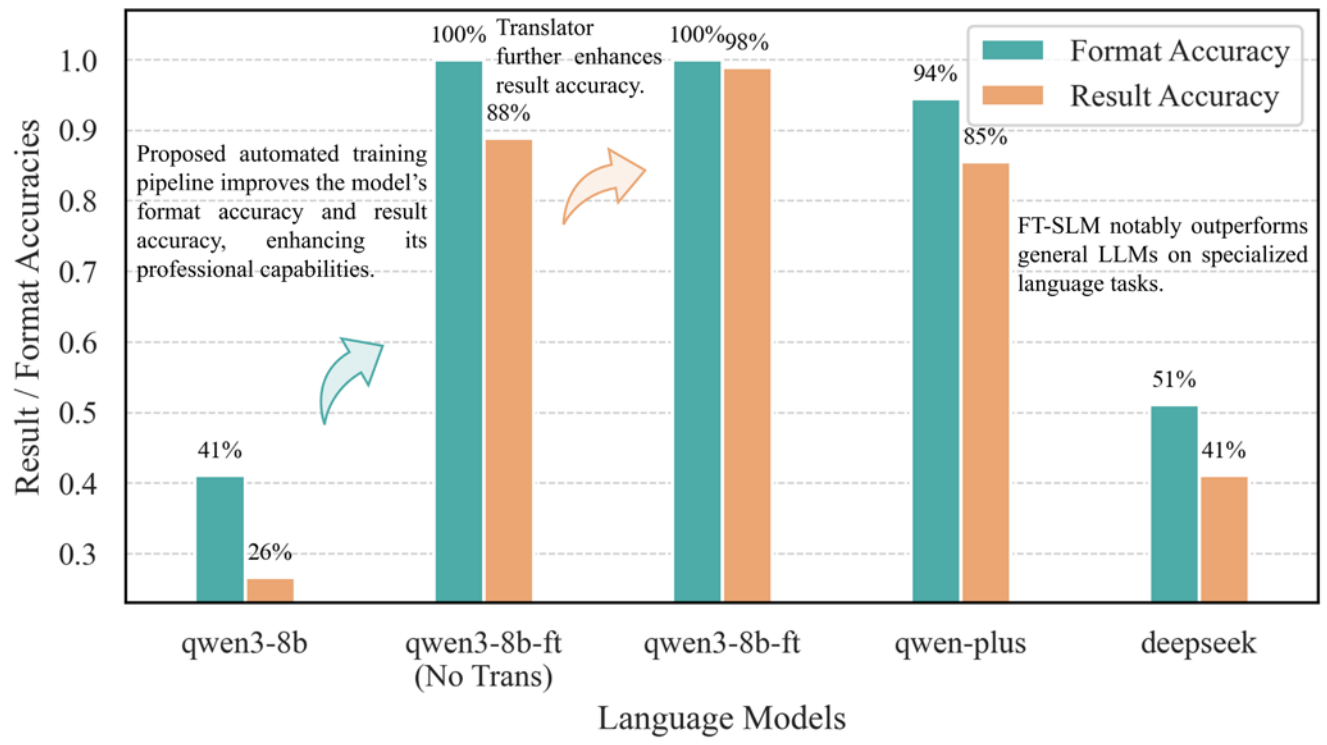

Fig. 4. Performance of different language models on the subtask of adjusting ADN model.

On the other hand, comparison with the two general LLMs further reveals that qwen3-8b-ft achieves substantially higher format accuracy and result accuracy. Although general LLMs exhibit stronger generalization capabilities across diverse domains, the FT-SLM demonstrates superior performance on this specific subtask. This observation provides additional evidence for both the necessity of fine-tuning and the effectiveness of our proposed approach.

### C. *Performance of ADN-Agent and Baseline Methods*

In this subsection, we present and compare the performance of ADN-Agent against several baseline methods in Table III. As can be seen, among all methods, ADN-Agent achieves the best performance, attaining a result accuracy of 95.8%, which demonstrates the effectiveness of the proposed method for ADN operation requests. During experimentation, we observed that the primary sources of error in ADN-Agent stem from internal inaccuracies of the underlying general LLMs, such as content generation errors in qwen-plus, which achieves a 98.3% completion rate but occasionally produces incorrect outputs, and intent misinterpretation in deepseek, which exhibits a DSM usage accuracy of 97.5%. These issues are mainly intrinsic to the LLMs themselves and are not attributable to the proposed ADN-Agent architecture.

TABLE III
PERFORMANCE OF DIFFERENT METHODS
(A) PERFORMANCE OF DIFFERENT METHODS ON QWEN-PLUS

| Method | Completion Rate | DSM Usage Accuracy | Result Accuracy |
|---|---|---|---|
| ADN-Agent | 98.3% | 99.2% | 95.8% |
| Function-Call | 94.2% | 85.8% | 65.0% |
| Multi-LLM | 31.7% | 77.5% | 15.0% |
| No-Trans | 95.8% | 100% | 83.3% |
| No-FT | 98.3% | 98.3% | 89.2% |
| Zero-Shot | 90.0% | 70.8% | 50.0% |

(B) PERFORMANCE OF DIFFERENT METHODS ON DEEPSEEK

| Method | Completion Rate | DSM Accuracy | Result Accuracy |
|---|---|---|---|
| ADN-Agent | 100% | 97.5% | 95.8% |
| Function-Call | 89.2% | 65.8% | 66.7% |
| Multi-LLM | 59.2% | 80.8% | 33.3% |
| No-Trans | 96.7% | 98.3% | 83.3% |
| No-FT | 83.3% | 97.5% | 65.0% |
| Zero-Shot | 72.5% | 50.8% | 44.2% |

For the baseline method Function-Call, although it achieves a high completion rate, its DSM usage accuracy drops significantly, ultimately reducing the result accuracy to approximately 65%. The most common error in the Function-Call paradigm is parameter mis-specification, which leads to incorrect DSM invocations and, consequently, erroneous final outputs. Since all function-calling commands in Function-Call are generated by a single LLM, this imposes a substantial cognitive burden on this LLM. As the number of DSMs increases, generating precise invocation commands for all DSMs becomes increasingly challenging. This phenomenon underscores the rationality of the proposed ADN-Agent architecture and its communication mechanism, where detailed commands are generated by different Translator modules. As for the multi-LLM collaboration paradigm, although it has been proven to be effective in some single-task settings, its performance suffers a significant degradation in the context of ADN operation, which involves multiple scenarios and diverse objectives. Since the multi-LLM collaboration method is provided only with raw implementation code rather than pre-encapsulated DSMs, its completion rate is very low, and the generated programs are often non-executable. This outcome highlights an inherent limitation of LLMs in synthesizing complex, functional systems from unstructured code alone, whereas DSMs effectively extend the capability boundaries of existing LLMs. On the other hand, due to the absence of an upper-level Planner for coordinated orchestration, the implementation of DSMs becomes less principled. Consequently, the multi-LLM collaboration paradigm achieves the poorest result accuracy among all evaluated methods.

As for the scalability, just as we mentioned earlier, the modular design significantly simplifies DSM updates within the ADN-Agent. When a DSM requires modification or a new one is introduced, only the specific Translator and Worker need adjustment, and the upper-level Planner remains unaffected. This stands in sharp contrast to Function-Call and multi-LLM collaboration approaches, which lack such modularity and consequently necessitate extensive modifications to the overall architecture and prompts for even minor updates.

Furthermore, our design offers superior scalability regarding the number of supported DSMs. Since the ADN-Agent's Planner relies on high-level functional descriptions rather than detailed implementation logic, it maintains a significantly shorter prompt length compared to baseline methods. This efficiency allows for the integration of a larger number of DSMs.

Based on the above observations, we summarize the respective advantages and limitations of these three DSM coordination paradigms in Table IV. The multi-dimensional comparison further demonstrates that, for future ADN operation tasks, the proposed ADN-Agent constitutes a more rational and scalable architectural design.

TABLE IV
MULTI-DIMENSIONAL COMPARISONS AMONG DIFFERENT DSM COORDINATION PARADIGMS

| Method | Description | Performance | Advantages | Disadvantages |
|---|---|---|---|---|
| ADN-Agent | The Planner recognizes the ADN operator's intent and decomposes the request into subtasks described in natural language. Translator module within each DSM translates these subtasks and generates executable commands. Worker module in the DSM then executes the commands to fulfill the subtasks. Finally, the Summarizer aggregates all partial results into a unified output. | ADN-Agent achieves the highest completion rate, DSM usage rate, and result accuracy among all of the methods. The predominant errors of ADN-Agent primarily stem from internal general LLM failures, such as misinterpretation of user intent and incorrect content generation. | 1. The burden on the Planner is substantially alleviated, enabling it to focus exclusively on intent recognition and task decomposition.<br>2. The communication mechanism is robust, reducing errors during DSM usage.<br>3. It exhibits high scalability and can efficiently integrate new DSMs. | 1. The initial design of the Planner and the Translator modules within DSMs requires both time and expert involvement.<br>2. The introduction of Translator modules in DSMs increases inference latency and token consumption. |
| Function-Call | All DSMs are encapsulated as callable functions and provided to a single LLM. During inference, this LLM can augment its content generation by invoking these functions, specifying their names, and filling corresponding parameters.<br>**Key difference from ADN-Agent:** The detailed DSM invocation commands are generated by the upper-level LLM itself rather than by the Translator modules. | Both DSM usage accuracy and result accuracy exhibit a significant decline, primarily due to function invocation errors and incorrect parameter specifications during function calls.<br>**Primary causes of errors:** Relying on a single upper-level LLM to handle all function invocations and command generation is prone to information loss and hallucination. | 1. During LLM inference, DSMs can be invoked directly in a simple and efficient manner, enabling a single-round interaction with one LLM to complete the entire task. | 1. As the number of DSMs increases, the cognitive load on the LLM rises significantly, thereby substantially increasing the probability of DSM invocation errors. |
| Multi-LLM | The implementation code of DSMs is provided to multiple LLMs, which act in a collaborative manner to generate the final complete executable program.<br>**Key difference from ADN-Agent:** The approach lacks an upper-level Planner for coordinated orchestration. Also, it does not provide pre-built DSMs for LLMs' direct invocation. | Both completion rate and result accuracy suffer a substantial decline, primarily due to the generation of infeasible programs and redundant or incorrect DSM implementations.<br>**Primary causes of errors:** A single infeasible code block can render the entire program non-executable. Moreover, the absence of a Planner for coordination leads to disordered content generation and execution sequencing among the LLMs. | 1. It eliminates the need to encapsulate DSMs and leverages the strong programming capabilities of existing LLMs to generate flexible and diverse programs. | 1. Reliable code generation becomes challenging under the complex multi-scenario and multi-objective operation settings. |

For the ablation baselines No-Trans and No-FT, both achieve a high completion rate and DSM usage accuracy. However, their result accuracy declines notably, particularly on operation requests that require the use of model adjustment FT-SLM. As demonstrated in our earlier experiments, the critical roles of the Translator module and fine-tuning are once again validated, especially with respect to their contribution to the overall efficacy of the ADN-Agent system. Finally, the Zero-Shot method exhibits performance degradation across all metrics, further highlighting the necessity of providing LLMs with few-shot examples to guide task execution. Since few-shot examples implicitly encode substantial information and domain knowledge, and given that imitation learning is one of the core strengths of existing general LLMs, providing a small set of few-shot examples is essential for correct content generation.

### *D. Discussions*

In this subsection, we delve into several common issues concerning the practical application of LLMs. First, we focus on the computational overhead, including inference latency and token consumption. To quantify this, we measure the average inference latency of the proposed ADN-Agent, Function-Call, and multi-LLM collaboration across different types of requests. The tested results are listed in Table V.

As can be observed from the presented results, Function-Call exhibits the lowest inference latency. This efficiency is attributed to its mechanism of encapsulating all DSMs as callable functions, which allows for immediate invocation upon receiving operation requests. In contrast, ADN-Agent incurs a higher processing time owing to the introduced Translator module, which is also one of the limitations of ADN-Agent. As for multi-LLM collaboration, since there are no encapsulated DSMs available, it necessitates coordination among various LLMs to complete the corresponding tasks. Consequently, it has the highest inference latency.

TABLE V
AVERAGE INFERENCE LATENCY OF DIFFERENT METHODS
(A) AVERAGE INFERENCE LATENCY ON QWEN-PLUS

| Method | Situation Awareness | Decision-Making | Operation Analysis |
|---|---|---|---|
| ADN-Agent | 7.82s | 18.27s | 40.70s |
| Function-Call | 4.57s | 15.09s | 37.57s |
| Multi-LLM | 31.16s | 89.18s | 93.00s |

(B) AVERAGE INFERENCE LATENCY ON DEEPSEEK

| Method | Situation Awareness | Decision-Making | Operation Analysis |
|---|---|---|---|
| ADN-Agent | 15.58s | 25.23s | 46.78s |
| Function-Call | 12.98s | 23.00s | 44.87s |
| Multi-LLM | 118.92s | 146.15s | 174.60s |

As for the inference latency comparison between different types of requests, it is evident that situation awareness is the simplest among all categories. It typically requires only basic data and model processing, along with possible power flow calculations, resulting in the shortest duration. In contrast, decision-making requests involve modeling optimization problems and invoking commercial solvers, leading to a significant increase in latency compared to situation awareness. Finally, operation analysis incurs the highest latency of all types, as it entails multi-step analysis involving calls to the FT-SLM.

However, it is worth noting that the operational time interval for current ADN is approximately 15 minutes, which means that measurements are collected or control commands are issued every 15 minutes. In regions with less developed communication infrastructure, this interval can extend to an hour or even longer. Therefore, this inference latency introduced by LLMs is acceptable and will not cause perceptible delay or adverse impact on actual operations. On the other hand, our empirical observations indicate that manual execution of the entire process usually requires 10~30 minutes. This substantial difference further demonstrates the necessity and effectiveness of utilizing LLMs to alleviate the burden of ADN operators, achieving full automation while significantly reducing the required time.

Regarding token consumption, we test the average prompt token consumption and completion token consumption for each module within the proposed ADN-Agent. The results are listed as follows:

TABLE VI
AVERAGE TOKEN CONSUMPTION OF DIFFERENT MODULES

| Module | Prompt Tokens | Completion Tokens |
|---|---|---|
| Planner | 1052.9 | 186.3 |
| Translator | 245.7 | 10.8 |
| Summarizer | 470.7 | 37.4 |

Second, since LLM generation is based on natural language and involves a certain degree of randomness, whereas ADN operations have stringent safety requirements, it is important to develop an error-recovery mechanism, which encompasses how to prevent errors during content generation/DSM execution, how to detect them if they occur, and how to correct them once identified. To address this challenge, we believe it can be resolved through the following approaches:

1) The ADN-Agent architecture and the novel communication mechanism we propose actually possess a certain degree of inherent robustness. In the proposed ADN-Agent, the LLM related modules are primarily responsible for intent comprehension, task decomposition, instruction generation, and result summarization, rather than directly participating in mathematical optimization or physical calculations. These precise computations are performed by the specialized Workers within the DSM. This design effectively combines the complementary strengths of both LLMs and DSMs, avoiding total reliance on LLMs for all tasks, thereby minimizing the occurrence of errors related to mathematical computations and physical constraints. In addition, the input and output of the Translators serve as the intermediate states of the entire ADN-Agent execution. In the form of natural language, they assist the ADN operator in understanding the task execution status and intermediate processes, allowing for timely intervention in case of any possible errors.

2) Building upon the architecture, achieving effective error recovery first requires efficient error detection. For the DSM, its inputs and outputs possess clear physical meanings. When an error occurs, such as power generation exceeding built-in limitations, it will trigger the internal exception, alerting the Planner or Translator to corresponding errors. Regarding the content generation of LLMs, we also try to minimize the probability of errors by enforcing a standardized output format for every response, i.e., content must be decorated with specific tags and formatted in JSON. The advantage of this approach is that if the LLM encounters an internal error or hallucination during content generation, it is likely to fail in adhering to the agreed-upon format. We subsequently extract the content using regular expressions and parse the JSON. If this process fails, we deem the generated content invalid and prompt the LLM to regenerate the response.

3) Following error detection, the next step is error recovery. During the regeneration process, we also feed the error information detected from the previous attempt back to the corresponding LLM, such as constraint-violating parameters, incorrect output format, or syntax errors in functions. This allows the LLM to avoid repeating the same mistake in subsequent dialogues. Leveraging the powerful in-context learning capabilities of current LLMs, such an error-recovery mechanism can effectively rectify the generated error, thereby ensuring the robustness and reliability of the system. Specifically, for Translators, given their pivotal role as a bridge within the ADN-Agent architecture, we provide them with not only the information of the corresponding DSM but also the subtasks decomposed by the Planner and the original request from the ADN operator. This multifaceted information allows for mutual verification, thereby preventing the accumulation and propagation of intermediate errors.

4) Ultimately, to ensure the highest level of reliability, operations concerning critical grid infrastructure must undergo a comprehensive validation phase. This involves a synchronized verification by the grid operating system alongside manual inspection by on-duty ADN operators. Far from being an optional add-on, this serves as a necessary prerequisite for execution, providing an ultimate guarantee of safety and preventing potential systemic risks.

## V. Conclusion

To facilitate efficient coordination among DSMs, this paper introduces the ADN-Agent framework, a dedicated communication mechanism, and an automated training pipeline for FT-SLMs. These integrated components collectively enhance the capability and adaptability of LLMs within complex, multi-scenario, and multi-objective ADN operational environments. Comprehensive comparative analyses and ablation studies validate the effectiveness of our approach, demonstrating that ADN-Agent constitutes a promising architectural paradigm for the future of ADN operations.

Looking ahead, a challenging yet valuable research direction involves investigating how ADN-Agent can augment its capabilities through evolvable self-learning. Furthermore, we aim to explore the adaptive creation of new DSMs to dynamically accommodate emerging requirements, thereby ensuring the system's long-term scalability and resilience.

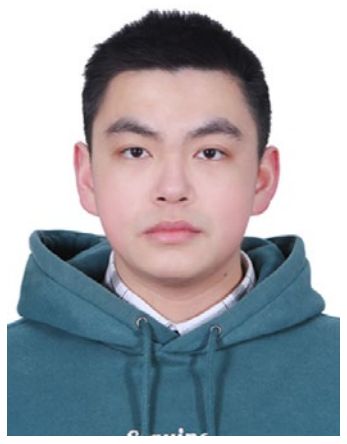

**Xu Yang** received the B.S. degree from the Department of Electrical Engineering, Tsinghua University, Beijing, China, in 2022, where he is currently pursuing the Ph.D. degree. His research interests include active distribution system operation and control, safe and robust reinforcement learning, large language model, and their applications in the energy system.

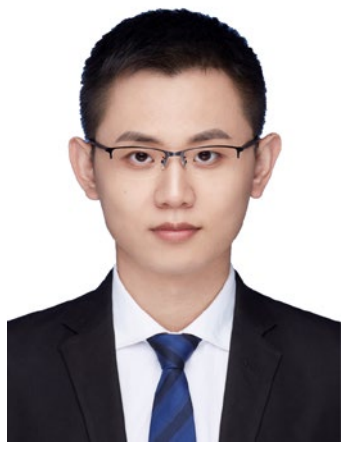

**Chenhui Lin** received the B.S. and Ph.D. degrees from the Department of Electrical Engineering, Tsinghua University, Beijing, China, in 2015 and 2020, respectively. He is currently a Research Assistant Professor with the Department of Electrical Engineering, Tsinghua University. His research interests include the operation and control of power systems integrated with renewable energy.

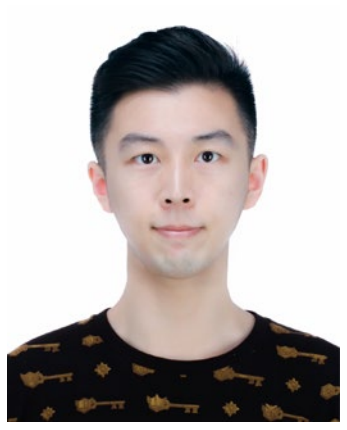

**Haotian Liu** received the B.S. and Ph.D. degrees from the Department of Electrical Engineering, Tsinghua University, Beijing, China, in 2018 and 2023, respectively. He is currently a Postdoctor with Tsinghua University. His research interests include active distribution system operation and control, model-free control, machine learning especially reinforcement learning, and their applications in the energy system.

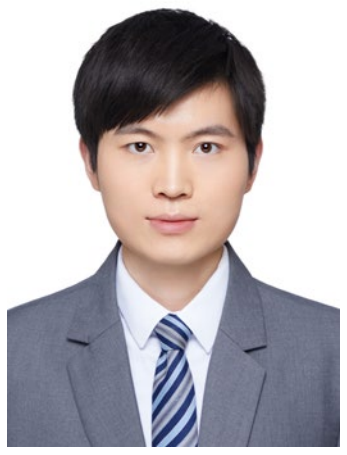

**Qi Wang** received the B.S. degree in 2019 from the School of Electrical Engineering and Automation at Harbin Institute of Technology and the Ph.D. degree in 2024 from the Department of Electrical Engineering at Tsinghua University. He is currently a Postdoc Fellow at The Hong Kong Polytechnic University. His research interests focus on efficient algorithms empowered by optimization and AI for energy system operation and control. Dr. Wang serves as a Youth Editorial Board Member of Protection and Control of Modern Power Systems. He was awarded the Outstanding Doctoral Dissertation Award.

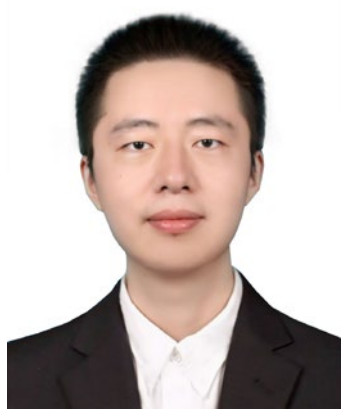

**Yue Yang** received the B.S. and Ph.D. degrees from the Department of Electrical Engineering, Tsinghua University, Beijing, China, in 2017 and 2022, respectively. He is currently an Assistant Professor with the Hefei University of Technology, Hefei, China. His research interests include optimization and control in power system with integration of renewable energy.

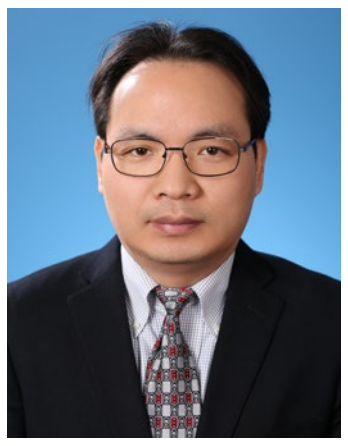

**Wenchuan Wu** received the B.S., M.S., and Ph.D. degrees from the Department of Electrical Engineering, Tsinghua University, Beijing, China, in 1996, 1998, and 2003, respectively. He is currently a Full Professor with Tsinghua University. His research interests include energy management system, active distribution system operation and control, machine learning and its application in energy system.